\documentclass[twocolumn,showpacs,aps,pra,amsmath,amssymb,superscriptaddress]{revtex4-1}
\usepackage{bm,color,bbm}
\usepackage{hyperref,mathtools,graphicx,natbib}

\newcommand{\ket}[1]{|{#1}\rangle}
\newcommand{\bra}[1]{\langle{#1}|}
\newcommand{\braket}[2]{\langle {#1} | {#2} \rangle}

\usepackage{times}
\newcommand{\beq}{\begin{equation}}
\newcommand{\eeq}{\end{equation}}
\newcommand{\bqa}{\begin{eqnarray}}
\newcommand{\eqa}{\end{eqnarray}}
\newcommand{\nn}{\nonumber}

\newcommand{\smallfrac}[2]{\mbox{$\frac{#1}{#2}$}}

\newcommand{\half}{\smallfrac{1}{2}}

\newcommand{\sq}[1]{\left[ {#1} \right]}

\newcommand{\tr}[1]{{\rm Tr}\sq{ {#1} }}

\newcommand{\blk}{\color{black}}

\definecolor{maroon}{rgb}{0.7,0,0}

\definecolor{ngreen}{rgb}{0.3,0.7,0.3}

\definecolor{golden}{rgb}{0.8,0.6,0.1}

\begin{document}
%\title{Products of weak values: uncertainty relations, optimal estimates, complementarity and incompatibility}
\title{Products of weak values: uncertainty relations, complementarity and incompatibility}
%\title{Uncertainty, complementarity and incompatibility relations with products of weak values }

\author{Michael J. W. Hall}

\affiliation{Centre for Quantum Computation and Communication Technology (Australian Research Council), Centre for Quantum Dynamics, Griffith University, Brisbane, QLD 4111, Australia}

\author{Arun Kumar Pati}

\email{akpati@hri.res.in}

\affiliation{Quantum Information and Computation Group,\\
Harish-Chandra Research Institute, Chhatnag Road, Jhunsi,
Allahabad 211 019, India}

\affiliation{Department of Mathematics,
Zhejiang University, Hangzhou 310027, PR~ China}

\author{Junde Wu}
\affiliation{Department of Mathematics,
Zhejiang University, Hangzhou 310027, PR~ China}

%\author{}
%\email{}

%\affiliation{}

%\date{\today}

\begin{abstract}
The products of weak values of quantum observables are shown to be of value in deriving quantum uncertainty and complementarity relations, for both weak and 
strong measurement statistics.  First, a `product representation formula'   allows the standard Heisenberg uncertainty relation to be derived from a 
classical uncertainty relation for complex random variables. We show this formula also leads to strong uncertainty relations for unitary operators, and underlies an 
interpretation of weak values as optimal (complex) estimates of quantum observables.  Furthermore, we show that  two incompatible observables that are weakly and strongly measured in a weak measurement 
context  obey a
complementarity relation under the interchange of these observables, in the form of an upper bound on the product of the corresponding weak values.  
Moreover, general tradeoff relations, between weak purity, quantum purity and quantum incompatibility, and also between weak and strong joint probability 
distributions, are obtained based on products of  real and imaginary components of weak values, where these relations quantify the degree to which weak 
probabilities can take anomalous values in a given context.
\end{abstract}


\maketitle

\section{ Introduction}
Quantum theory has many counter intuitive features such as wave-particle duality, interference, 
entanglement and non-locality, and 
these features make the subject 
exciting even after ninety years since its initial formulation. To this weird list, the weak value adds another twist 
making quantum theory even stranger than before.
The concept of weak value was introduced by Aharonov, Albert and Vaidman
\cite{aha,av} while investigating the properties of a quantum system in pre and post-selected ensembles.
 If a system is weakly coupled to an apparatus, then upon post-selection of the system state, 
 the apparatus pointer observable is shifted on average by (the real part of) a weak value \cite{average}.

The weak value can have strange properties. For example,  it is a complex number in general,
and its real part can take values outside the spectrum of the observable being measured. 
 This is in sharp contrast to the average value of an observable when measured by a strong coupling to an apparatus, which is always bounded by the smallest 
 and largest eigenvalues. It gives rise to 
 the notion of anomalous weak value for an observable \cite{aha,av}, recently sharpened in \cite{akp}.
The concepts of weak measurements and weak values have since been generalized in various directions 
\cite{sw1,ADL,sw2,nori,shik,rub}  
and have found numerous applications \cite{eric,sandu,cho,wise1,mir,js,ams1,ams2,pdav,jsl1,hof,jd,geno}.  However, while the separate real and imaginary components
of weak values have been given various interpretations in the literature \cite{joh1,hall,joh2,joz,dress,hof,dressint,inter}, the weak value itself as a complex number
has not.  This paper is in part concerned with redressing this issue, via the consideration of {\it products} of weak values.

It is well known that weak values respect sums, but not products.  That is, the weak value of the sum of two observables is just the sum of their weak values, 
whereas the weak value of the product of the two observables is not equal to the product of their weak values.  The latter feature has been suggested as 
underlying curious phenomena such as the quantum Chesire cat, where one property of a quantum system, such as its spin, appears to be spatially separated 
from another property, such as its position \cite{cat1,cat2,cat3}.  
Nevertheless, we show in this paper that the products of weak values are well worth investigating, leading to a direct connection between quantum and classical uncertainty relations, the interpretation of weak values as optimal estimates, a form of complementarity for weak measurement scenarios,  and tradeoffs between the incompatibility of quantum observables and the purity of their weak joint probability distribution.

We first review and generalise a beautiful representation theorem by Shikano and Hosoya in Sec.~II.A, that directly connects the average  of 
the product of two quantum observables with the corresponding average product of their corresponding weak values \cite{shik}. This  is of direct operational significance in allowing the quantum average of any product to be reconstructed from the weak values of the corresponding observables. In this sense weak values provide a hidden variable model for the averages of a given set of quantum observables 
and their pairwise products. In Sec.~II.B we show 
that the representation formula  provides a simple derivation of the Heisenberg uncertainty relation. In fact, the latter may be reinterpreted as a
classical dispersion relation for complex random variables.   We also use the representation formula to  obtain a strong uncertainty relation for unitary operators, with a simple geometric interpretation. 
In Sec.~II.C we further show that the representation formula leads naturally to the interpretation of a weak value as the optimal estimate of a given observable,
based on the measurement outcome of a second observable. This 
interpretation is of particular interest in that it makes no reference to the decomposition of the weak value into real and imaginary parts (nor to weak meaurements), unlike previous connections between weak values and estimates \cite{joh1,hall,dressint}.

We then consider various scenarios  in which products of weak values lead to physical implications independent of the product representation formula. 
Thus, in Sec.~III we obtain a complementarity relation for the weak values of two non-commuting projection operators, that restricts the degree to which they can take 
anomalous values outside the interval $[0,1]$. In particular, it is possible to weakly measure the projection $|a\rangle\langle a|$ with postselection on projection 
$|b\rangle\langle b|$, or vice versa. We show that the product of the corresponding weak values for these complementary scenarios is restricted to be a positive real 
number, no greater than $|\langle a|b\rangle|^2$, with equality for all pure initial states.

In Sec.~IV we obtain general tradeoff relations that connect the weak joint probability distribution of two observables \cite{dressint,weakprob} with their 
degree of incompatibility, and with the corresponding strong joint probability distribution. It is also shown that, despite weak probabilities having anomalous values, 
the corresponding `weak purity' cannot exceed unity. Finally, conclusions are given in Sec.~V.

\section{Uncertainty relations and optimal estimates from products of weak values}

\subsection{Representation of products}

The weak value of a Hermitian operator $A$, for the pre-selected state $|\psi\rangle$ and the post-selected state $|\phi\rangle$, is defined by \cite{aha,av}
\beq \label{awdef}
A_w(\phi|\psi):= \frac{\langle\phi|A|\psi\rangle}{\langle \phi|\psi \rangle} .
\eeq
The real part of $A_w(\phi|\psi)$ corresponds to the average value of a weak measurement carried out on $|\psi\rangle$, when postselected on the final 
state $|\phi\rangle$ \cite{average}, although it may also be measured via averages of suitable strong measurements \cite{joh2}.  The imaginary part may similarly be 
measured via suitable weak and/or strong measurements \cite{joz,joh2,inter}.  The real and imaginary parts can also be directly related to properties of optimal estimates of $A$ from strong 
measurements on state $|\psi\rangle$ \cite{joh1,hall,weston,dressint}, and to measurement-induced disturbance \cite{hof,dress}.  It follows, therefore, that weak values have operational 
significance independently of weak measurements {\it per se}.

More generally, the role of the postselected state $|\phi\rangle$ may be replaced by the outcome of measuring some maximal positive operator valued measure (POVM) $M\equiv \{|m\rangle\langle m|\}$, 
with $\sum_m |m\rangle\langle m|=\hat1$, leading to the expression
\beq \label{av}
A_w(m|\psi) := \frac{\langle m|A|\psi\rangle}{\langle m|\psi \rangle},
\eeq
for the weak value of $A$ postselected on measurement outcome $M=m$.  Since the probability of measurement outcome $m$ on state $|\psi\rangle$ is
$p(m|\psi)=|\langle m|\psi \rangle|^2$, it immediately follows that the average of the weak value, over all possible outcomes, is given by~\cite{aha,av}
\beq \label{avsum}
\langle A_w\rangle_p := \sum_m p(m|\psi)\,A_w(m|\psi) =\langle \psi|A|\psi\rangle =:\langle A\rangle_\psi,
\eeq
i.e, by the average value of $A$ for state $|\psi\rangle$. Note that this reconstruction formula holds for any 
maximal measurement $M$, and also applies to non-Hermitian operators $A$ via linearity.

Remarkably, the above reconstruction formula for the average of an observable may be extended to a similar formula for products \cite{shik}
(generalised here to arbitrary non-Hermitian operators):
\[
 \sum_m p(m|\psi)\,A_w(m|\psi)^*\,B_w(m|\psi) =  \langle \psi|A^\dagger B|\psi\rangle 
\]
i.e.,
\beq \label{prod}
\langle A_w^*B_w\rangle_p = \langle A^\dagger B\rangle_\psi,
\eeq
as is readily checked by direct substitution.  Thus, the average of an operator product, with respect to a quantum state $|\psi\rangle$, can be replaced by the 
average of a product of weak values, with respect to the classical probability distribution $p(m|\psi)$. Note that this equation can alternatively be written as a 
relation between quantum and classical inner products, $(A_w,B_w)=(A,B)$. Equation~(\ref{prod}) reduces to Eq.~(\ref{avsum}) when $A=A^\dagger$ and $B=\hat 1$. 

The product representation formula has a clear operational significance. For example, if the weak values of two Hermitian operators $A$ and $B$,  postselected on
measurement $M$, are determined experimentally, then one can immediately recover not only the averages of the observables $A$ and $B$, but also the averages
of $A^2$, $B^2$, and $AB$---and, hence, the variances and covariances of $A$ and $B$ \cite{shik}.  One can also experimentally recover the average of the
operator $(A-B)^2$ from the weak values of $A$ and $B$, where this average appears in various error-disturbance and joint-measurement uncertainty relations
\cite{mo1,hall, weston,bra,cbra,ozmixed}. 

We  note from Eqs.~(\ref{avsum}) and (\ref{prod})  that weak values also provide a (complex) hidden variable model for the averages of a given set of 
quantum observables and their pairwise products. For example, the average values of all linear and quadratic functions of the annihilation and creation operators 
$a$ and $a^\dagger$ of a single mode field, including the quadrature observable $X_\theta=a e^{i\theta}+a^\dagger e^{-i\theta}$ and the number operator $a^\dagger a$, can be 
modelled for any state $|\psi\rangle$, via the corresponding weak values $a_w(m|\psi)$ and $a^\dagger_w(m|\psi)$ and classical probability distribution $p(m|\psi)$
(and for any choice of POVM $M\equiv \{|m\rangle\langle m|\}$). 
While it is possible to generalise the product representation formula and such hidden variable models to arbitrary density operators \cite{density}, we note that it does not appear possible to extend them to products of three or more observables.  \blk

We give two particular applications of the product representation formula in the following subsections.

\subsection{Uncertainty relations from weak values}

Complex random variables are standard tools in classical signal processing and information theory \cite{sig1, sig2}.  
A complex random variable $\alpha=\alpha_1+i\alpha_2$ is described by some real and positive probability density $p(\alpha)$. The expectation value of function $f(\alpha)$ is 
then given by $\langle f(\alpha)\rangle:=\int d\alpha \,p(\alpha)f(\alpha)$, where the integral is over the complex plane with respect to the uniform measure. If $\alpha$ is restricted to some set of discrete values, $\{\alpha_j\}$, this reduces to the form $\langle f(\alpha)\rangle=\sum_j p_j\,\alpha_j$ for a corresponding discrete probability distribution $\{p_j\}$. 

A well known 
example in quantum mechanics is the outcome, $\alpha$, of a balanced homodyne measurement on a single-mode field. This measurement is described by the coherent 
state POVM  $\{\pi^{-1}|\alpha\rangle\langle\alpha|\}$, with corresponding probability density for the field state $\rho$ given by the 
Husimi Q-function $p(\alpha)=\pi^{-1}\langle\alpha|\rho|\alpha\rangle$ \cite{qfunction}. 

The variance of a complex random variable $\alpha$ is just the average mean square distance between $\alpha$ and its mean value \cite{sig1,sig2}, i.e.,
\beq \label{vardef}
{\rm Var} \,\alpha := \langle \,\left|\alpha-\langle \alpha\rangle\right|^2\rangle  = \langle \,|\alpha|^2\rangle - |\langle \alpha\rangle|^2 .
\eeq
Similarly, the covariance of two such random variables, $\alpha$ and $\beta$, with respect to a joint probability distribution $p(\alpha,\beta)$, is defined by
\beq \label{cov}
{\rm Cov}(\alpha,\beta) := \langle \,(\alpha-\langle \alpha\rangle)^*(\beta-\langle \beta\rangle)\rangle  = \langle \,\alpha^*\beta\rangle - \langle \alpha^*\rangle\langle \beta\rangle 
\eeq
with $\langle f(\alpha,\beta)\rangle:=\int d\alpha d\beta\,p(\alpha,\beta)\,f(\alpha,\beta)$. 
Thus, ${\rm Var}\,\alpha={\rm Cov}(\alpha,\alpha)$, and one immediately has the {\it classical} uncertainty relation \cite{sig2}
\beq \label{schwarz}
{\rm Var}\,\alpha\,{\rm Var}\,\beta \geq \left| {\rm Cov}(\alpha,\beta) \right|^2
\eeq
from the Schwarz inequality for complex numbers (or by noting that the determinant of the nonnegative $2\times2$ matrix $\langle zz^\dagger\rangle$ must be positive, for $z:=(\alpha-\langle\alpha\rangle,\beta-\langle\beta\rangle)$). \blk
Choosing $\alpha=A_w(m|\psi)$, $\beta=B_w(m|\psi)$,  and probability distribution $p(m|\psi)$, this  classical uncertainty relation reduces to 
\beq \label{weakuncert}
{\rm Var}_p A_w\,{\rm Var}_p B_w \geq \left| {\rm Cov}_p(A_w,B_w) \right|^2 .
\eeq
for the case of weak values. We now apply this relation in two scenarios of fundamental interest.

\subsubsection{Heisenberg inequality as a classical uncertainty relation}

The standard Heisenberg uncertainty relation \cite{heis,ken,rob} for two non-commuting Hermitian observables $A$ and $B$ follows directly from the product representation 
formula in Eq.~(\ref{prod}) and the classical uncertainty relation in Eq.~(\ref{weakuncert}). First, note from the definition of variance for classical random variables in Eq.~(\ref{vardef}) that 
\begin{align*}
	{\rm Var}_p A_w &= \langle\, |A_w|^2\rangle_p - |\langle A_w\rangle_p|^2\\
	&= \langle A^\dagger A\rangle_\psi - |\langle A\rangle_\psi|^2 = {\rm Var}_\psi A,
\end{align*}
where the second line follows immediately from Eqs.~(\ref{avsum}) and Eq.~(\ref{prod}). Similarly, ${\rm Var}_p B_w={\rm Var}_\psi A$, while from the definition of classical covariance in Eq.~(\ref{cov}) one has
\begin{align*}
	{\rm Cov}_p(A_w,B_w) &=  \langle A_w^*B_w\rangle_p - \langle A_w^*\rangle_p\langle B_w\rangle_p\\
	&= \langle A^\dagger B\rangle_\psi -\langle A\rangle_\psi^*\langle B\rangle_\psi\\ 
	&= \langle AB\rangle_\psi -\langle A\rangle_\psi \langle B\rangle_\psi ,
\end{align*}
again using Eq.~(\ref{avsum}) and the product representation formula (\ref{prod}).
Substituting these expressions into the classical uncertainty relation ~(\ref{weakuncert}), and decomposing the right hand side into real and imaginary parts, then yields 
\begin{align} \label{uncert}
{\rm Var}_\psi A\, {\rm Var}_\psi B &\geq  \left| \langle AB\rangle_\psi - \langle A\rangle_\psi \langle B\rangle_\psi\right|^2  \nn\\
&= {\rm Cov}_\psi(A,B)^2 + \frac{1}{4}\left|\langle [A,B]\rangle_\psi\right|^2 ,
\end{align}
with the quantum covariance defined by ${\rm Cov}_\psi(A,B):=\half\langle  AB+BA\rangle_\psi - \langle A\rangle_\psi \langle B\rangle_\psi$. 
This may be recognised as Schr\"odinger's strengthened form of the Heisenberg uncertainty relation \cite{schr}.
Thus, the standard quantum uncertainty relation may be reinterpreted as a {\it classical} uncertainty relation for weak values. 

It is a curious fact that one of the fundamental relations of quantum mechanics, namely, the Heisenberg uncertainty relation, can be 
understood as a classical uncertainty relation for complex random variables. In this sense the weak value approach removes the mystery associated with this
uncertainty relation.
Recently, quantum uncertainty relations have been proved \cite{lorenzo} which go beyond the Robertson-Schr{\"o}dinger uncertainty 
relation. It may be interesting to explore if one can view these also as classical uncertainty relations for weak values.

\subsubsection{Uncertainty relations for unitary operators}

It is straightforward to also obtain useful uncertainty relations for non-Hermitian operators. First, note that the variance of a general operator is defined by \cite{levy}
\beq \label{vargen}
{\rm Var}_\psi A := \langle A^\dagger A\rangle_\psi -|\langle A\rangle_\psi|^2 = {\rm Var}_p A_w,
\eeq
where the second equality follows from the product representation formula (\ref{prod}). This quantity has the desirable properties of vanishing if and only if $|\psi\rangle$ is an eigenstate of $A$, and of reducing to the usual variance in the Hermitian case $A=A^\dagger$.  Thus, for example, recalling that the annihilation and creation operators $a$ and $a^\dagger$ of a single-mode bosonic field satisfy $[a,a^\dagger]=1$, their variances are related by
\beq 
{\rm Var}\,a^\dagger = \langle aa^\dagger\rangle - |\langle a^\dagger\rangle|^2 =  \langle a^\dagger a+1\rangle - |\langle a\rangle|^2 = {\rm Var}\,a +1 \geq 1 \blk
\eeq
implying immediately that $a^\dagger$ has no eigenstates.  

The classical uncertainty relation in Eq.~(\ref{weakuncert}) yields the generalisation
\begin{align} \label{uncertgen}
{\rm Var}_\psi A\, {\rm Var}_\psi B &\geq \left| {\rm Cov}_p(A_w,B_w) \right|^2 \nn\\
&= \left| \langle A^\dagger B\rangle_\psi - \langle A^\dagger \rangle_\psi \langle B\rangle_\psi\right|^2 
\end{align}
of Eq.~(\ref{uncert}), to general operators $A$ and $B$, where the second line follows via the product representation formula~(\ref{prod}) and the definition in Eq.~(\ref{cov}).
This quantum uncertainty relation is equivalent to Eq.~(8) of Pati {\it et al.} \cite{pati}, and again is seen to be equivalent to a classical uncertainty relation for weak values.

As an application of the generalised uncertainty relation~(\ref{uncertgen}), consider the case of two unitary operators $U$ and $V$. Defining 
$$u:=|\langle U\rangle_\psi|,\qquad v:=|\langle V\rangle_\psi|,$$
the variance of $U$ is given by $1-u^2$, in agreement with previous proposals in the literature \cite{levy,pati,massar,patiun}, and substitution into Eq.~(\ref{uncertgen}) yields
\begin{align} \label{uv}
(1-u^2) \,(1-v^2) &\geq \left|\langle U^\dagger V\rangle_\psi -\langle U^\dagger \rangle_\psi\langle V\rangle_\psi \right|^2 \nn
\\
&\geq \left| |\langle U^\dagger V\rangle_\psi| - uv\right|^2 , 
\end{align}
which may be rewritten as the uncertainty relation
\beq \label{unitary}
u^2 +v^2 -2uv |\langle U^\dagger V\rangle_\psi|  \leq 1-|\langle U^\dagger V\rangle_\psi|^2.
\eeq
for two unitary operators.  Note that $|\langle U^\dagger V\rangle_\psi|^2$ is the overlap of the states $U|\psi\rangle$ and $V|\psi\rangle$.  Hence, the overlap plays a role analogous to the commutator in the Heisenberg uncertainty relation.

Zero uncertainties for $U$ and $V$ correspond to $u=v=1$. Noting that the weaker uncertainty relation
\beq \label{hyp}
u v\leq \frac{1 + |\langle U^\dagger V\rangle_\psi|}{2}
\eeq
follows directly from Eq.~(\ref{unitary}), using the inequality $2uv \leq u^2+v^2$, it follows that zero uncertainties are possibly only in the case of a unit overlap, $|\langle U^\dagger V\rangle_\psi|^2=1$, as expected.

\begin{figure}[t]
	\centering
	\includegraphics[width=0.45\textwidth]{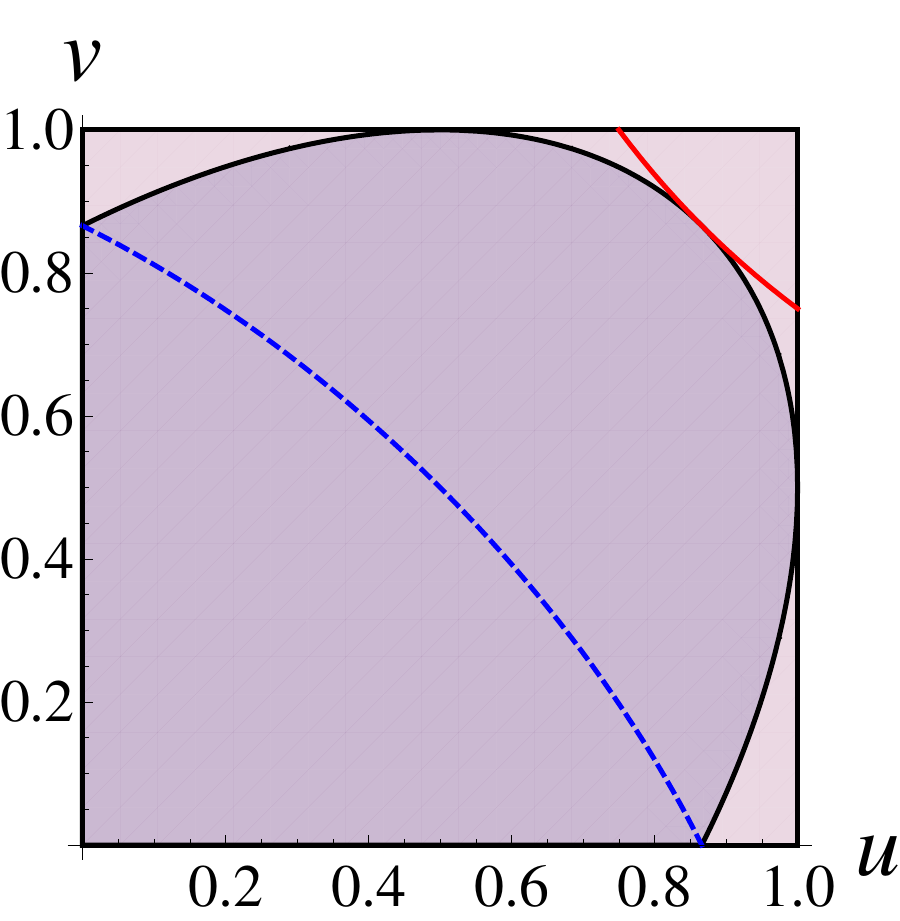}
	\caption{(color online). Uncertainty relations for unitary operators. The uncertainty relation in Eq.~(\ref{unitary}) is depicted for an overlap of 1/4, and restricts $u$ and $v$ to lie in the intersection of an ellipse with the positive quadrant (purple region). The corresponding weaker uncertainty relation in Eq.~(\ref{hyp}) is also shown, corresponding to the region  below a hyperbola that is tangent to the ellipse (solid red curve). Finally, the stronger uncertainty relation in Eq.~(\ref{imp}) is depicted for the case $\Phi=\pi$, and constrains $u$ and $v$ to lie in the elliptical region below the dashed blue curve. }
\end{figure}

More generally,   defining $x=(u+v)/\sqrt{2}$ and $y=(u-v)/\sqrt{2}$, Eq.~(\ref{unitary}) takes the form
\beq \label{ellipse}
\frac{x^2}{1+|\langle U^\dagger V\rangle_\psi|} + \frac{y^2}{1-|\langle U^\dagger V\rangle_\psi|} \leq 1
\eeq
Hence, the uncertainty relation constrains $u$ and $v$ to lie within an ellipse oriented diagonally in the $uv$-plane, determined by the overlap of  $U|\psi\rangle$ and $V|\psi\rangle$, as depicted in Fig.~1. Note that the area of this ellipse is proportional to the square root of $1-|\langle U^\dagger V\rangle_\psi|^2$, and hence the uncertainty relation becomes more constraining as the overlap increases. The weaker uncertainty relation in Eq.~(\ref{hyp}) is also depicted in Fig.~1.

Similar elliptical constraints for $u$ and $v$ have been obtained previously by Massar and Spindel for the special case $U V=e^{i\phi}VU$ \cite{massar} (see also \cite{rudnicki}).
A circular constraint for the general case has been recently obtained by Bagchi and Pati \cite{patiun}, which is typically tighter than 
that given in Ref.~\cite{massar}. We can obtain an even stronger constraint 
by proceeding directly from the first inequality in 
Eq.~(\ref{uv}) above, to obtain the stronger uncertainty relation (see also Eq.~(4) of Ref.~\cite{patiun})
\beq \label{imp}
u^2 +v^2 -2uv |\langle U^\dagger V\rangle_\psi|\cos\Phi  \leq 1-|\langle U^\dagger V\rangle_\psi|^2.
\eeq
Here $\Phi$ denotes the phase of the complex number $\langle U \rangle_\psi\langle U^\dagger V\rangle_\psi\langle V^\dagger\rangle_\psi$, where the latter is the 
Bargman invariant associated with $|\psi\rangle$, $U|\psi\rangle$ and $V|\psi\rangle$ \cite{patiun,barg}.  

Uncertainty relation (\ref{imp})  yields an 
ellipse in the $uv$-plane, similarly to the weaker relation in Eq.~(\ref{unitary}) (where the latter corresponds to $\Phi=0$), but bounds a smaller region in the positive $uv$-quadrant, as shown in Fig.~1. Note, 
however, that the stronger relation requires further knowledge in addition to the overlap of $U$ and $V$ for state $|\psi\rangle$.  This is analogous to the 
strengthened Heisenberg inequality in Eq.~(\ref{uncert}), which requires further knowledge in addition to the average commutator of $A$ and $B$ for state $|\psi\rangle$.

Finally, note that it is easy to extend the above uncertainty relations to any density operator $\rho$, by replacing $|\psi \rangle$ with a  purification of $\rho$ on a suitably 
extended Hilbert space.  This has the effect of replacing averages over $|\psi \rangle$ with averages over  $\rho$ in the above results.

\subsection{Weak values as optimal estimates}

We have shown that quantum uncertainty relations for the variances of Hermitian and non-Hermitian operators correspond to classical uncertainty relations for
weak values.  This suggests that the `quantumness' in these relations is modelled by the assignment of a complex value $A_w(m|\psi)$ to observable $A$, for 
measurement outcome $M=m$ on state $|\psi\rangle$.

Here we show that this can be given a more precise sense: the weak value $A_w(m)$ in Eq.~(\ref{av}) may be interpreted as the best possible estimate of $A$, 
from measurement outcome $M=m$ on state $|\psi\rangle$, provided that complex numbers are permitted as estimates (as is the case, for example, in balanced homodyne detection, where one estimates $a$ via the complex outcomes of the coherent state POVM $\{\pi^{-1}|\alpha\rangle\langle\alpha|\}$ \cite{qfunction}).

For simplicity we will restrict attention to the case that $M$ corresponds to a Hermitian operator.  Any estimate $A=\alpha_m$ from outcome $M=m$ then corresponds to 
measurement of $A^{\rm est}:=\sum_m \alpha_m\,|m\rangle\langle m|$. The weak value of $A^{\rm est}$ for outcome $m$ follows as $A^{\rm est}_w(m|\psi)=\alpha_m$, using Eq.~(\ref{av}).  Now, 
one possible measure of the degree to which $A^{\rm est}$ provides a `good' estimate of $A$ is the average value of $|A-A^{\rm est}|^2$ \cite{msd}.  Thus, we  define the mean 
square deviation of the estimate by
\begin{align}
\epsilon^2:= \langle\, |A-A^{\rm est}|^2\rangle_\psi &= \langle\, |A_w - A^{\rm est}_w|^2\rangle_p \nn\\
 &= \sum_m p(m|\psi) \,\left|(A_w(m|\psi)-\alpha_m)\right|^2,
\end{align}
where the  equality in the first line follows from the product representation formula in Eq.~(\ref{prod}).  This quantity is nonnegative, and clearly vanishes if and only if the optimal estimate
\beq
\alpha_m^{\rm opt} := A_w(m|\psi).
\eeq
is made.  Thus the best possible estimate of $A$ is its weak value, as claimed.

The above result is of interest in providing an interpretation of the weak value that does not rely on decomposing it into real and imaginary parts, and which is, moreover, independent of the concept of weak measurements. 
It may be regarded as a generalisation of the case where the optimal estimate is constrained to be a real number (in which case the optimal estimate becomes the real part of the weak value, with a mean square error related to the imaginary part \cite{joh1,hall,dressint}).

\section{Complementarity of weak values}

In quantum theory, complementarity imposes limitations on our ability to unambiguously define and measure aspects of quantum systems in a single measurement setup. Indeed, Bohr wrote that ``...it is only the mutual exclusion of any two experimental procedures, permitting the unambiguous definition of complementary physical quantities, which provides room for new physical laws, the coexistence of which might at first sight appear irreconcilable with the basic principles of science. It is just this entirely new situation as regards the description of physical phenomena, that the notion of {\it complementarity} aims at characterizing'' \cite{bohr}. \blk

In the context of weak measurements, it is possible that one can probe two
complementary aspects of a quantum system at some price (e.g. introducing noise), as the apparatus 
interacts with the system weakly, allowing
a gentle observation without disturbing the system too much \cite{geno}. However, it turns out that there is, nevertheless, 
a strong type of 
complementarity: between a given weak measurement procedure, and the mutually exclusive procedure obtained by interchanging the weak and the strong components thereof.  \blk

In particular, we consider the case of weak measurements involving  projection operators
\beq  
A^a=|a\rangle\langle a|,\qquad B^b=|b\rangle\langle b|,
\eeq 
corresponding to the eigenvalue decompositions of two nondegenerate observables $A=\sum_a aA^a$ and $B=\sum_b bB^b$. For a given initial state $|\psi\rangle$ there are then two complementary weak measurement scenarios: a weak measurement of projector $A^a$ postselected on state $|b\rangle$, i.e, on $B=b$, and a weak measurement of 
projector $B^b$ postselected on state $|a\rangle$, i.e., on $A=a$.  The corresponding weak values follow from either of Eqs.~(\ref{awdef}) and (\ref{av}) as
\beq \label{probcom}
A^a_w(b|\psi) = \frac{\langle b|a\rangle\langle a|\psi\rangle}{\langle b|\psi\rangle}, ~~~~
B^b_w(a|\psi) = \frac{\langle a|b\rangle\langle b|\psi\rangle}{\langle a|\psi\rangle}.
\eeq

We see that the weak values connect wavefunctions directly in complementary bases. For example, we have
\begin{align}
A^a_w(b|\psi)\, \psi(b)  &= \bra{b} a \rangle\, \psi(a), \nonumber\\
B^b_w(a|\psi)\, \psi(a) & = \bra{a} b \rangle \,\psi(b),
\end{align}
where $\psi(a)$ and $\psi(b)$ are the wavefunctions in the eigenbasis representations of $A$ and $B$, respectively.
The interesting point to note here is $\psi(a)$ and $\psi(b)$ are directly related without a unitary transformation: the weak values act as `filters' that connect two complementary aspects directly.

Next, we ask can these two weak values be arbitrarily large at the same time? Strangely, not.
First, note that the weak values for the projectors $A^a$ and $B^b$ can be expressed as
 the sum of the average of the projectors in the state $\ket{\psi}$ and an anomalous part \cite{akp}
 \begin{align}
\label{vaid}
A^a_w(b|\psi) &= \langle A^a \rangle_\psi  +  \Delta_\psi A^a \,\frac{ \braket{b}{ {\bar{\psi_a}}} }{ \braket{b}{\psi} }, \nonumber\\
B^b_w(a|\psi)  &= \langle B_b\rangle_\psi  +  \Delta_\psi B^b\, \frac{ \braket{a}{ {\bar{\psi_b}}} }{ \braket{a}{\psi} },
\end{align}
 where $\Delta_\psi A^a:=({\rm Var}_\psi A^a)^{1/2}$ is the uncertainty of the projector in the state $\ket{\psi}$, $\ket{\bar{\psi_a}}$ is a state orthogonal to $\ket{\psi}$, 
 and similar definitions 
hold for the other projector $\Pi_b$.
This shows that the weak values of these projectors can be large, and lie outside the eigenvalue range $[0,1]$ of the projectors.
However, both  weak values cannot be large at the same time.
Indeed, from Eq.~(\ref{probcom}) the product of these weak values satisfies \blk
\begin{align} \label{comp}
A^a_w(b|\psi)\,   B^b_w(a|\psi) = |\bra{a} b \rangle|^2  \le 1.
\end{align}
Thus, even though individually each of these weak values can be complex, with arbitrarily large moduluses, their product is real, independent of the
pre-selected state, and bounded by unity.
This represents a new kind of complementarity between the weak and strong components of quantum weak measurements. 

This type of complementarity also holds for the scenario of a weak momentum measurement postselected on the result of a strong position measurement, and its converse. In this case the corresponding 
weak values for state $|\psi\rangle$ are given by 
\beq 
P^p_w(x|\psi) = \frac{\langle x|p\rangle\langle p|\psi\rangle}{\langle x|\psi\rangle}, ~~~~
X^x_w(p|\psi) = \frac{\langle p|x\rangle\langle x|\psi\rangle}{\langle p|\psi\rangle},
\eeq
and it is easily checked that the product of these two weak values satisfy the condition 
\begin{align}
X^x_w(p|\psi) \, P^p_w(x|\psi)  = \frac{1}{2\pi\hbar}.
\end{align}
Thus, the complementarity of weak values of two non-commuting projectors is a general feature of quantum systems, 
that holds in both finite and infinite dimensions. 

Finally, we note that it is straightforward to generalise the above complementarity relations to density operators.  In particular, recalling that the weak value of observable $A$ for density operator $\rho$, postselected on POVM $M\equiv \{M_m\}$, is given by \cite{hall,dressint}
\beq \label{weakrho}
A_w(m|\rho) = \frac{\tr{\rho M_m A}}{\tr{\rho M_m}}, \blk
\eeq 
one has 
\begin{align} \label{comgen}
A^a_w(b|\rho) \, B^b_w(a|\rho) &= \frac{\langle b|\rho|a\rangle\langle a|b\rangle}{\langle b|\rho|b\rangle} ~
\frac{\langle a|\rho|b\rangle \langle b|a\rangle}{\langle a|\rho|a\rangle}
 \nn\\
&= |\langle a|b\rangle|^2 \,\frac{|\langle a|\rho|b\rangle^2}{\langle a|\rho|a\rangle \langle b|\rho|b\rangle}\nn\\
&\leq |\langle a|b\rangle|^2 \leq 1,
\end{align}
where the first inequality follows from the Schwarz inequality. Thus, the product is again a positive real number, and one obtains a simple generalisation of the complementarity relation in Eq.~(\ref{comp}).

In the next section, we make more 
precise the notion of complementarity, by making connections to weak probabilities, weak purity and incompatibility.

\section{Weak probabilities and Incompatibility}

In the previous sections we have adressed the significance and applications of products of complex weak values. Here we focus on products related to their real and imaginary components.  In particular, we give natural and logically independent definitions of the weak purity and the incompatibility of observables, in Secs.~IV.A and~IV.B, and then show that these satisfy a number of tradeoff relations, related to properties of quantum purities and strong probability distributions, in Secs.~IV.C and~IV.D.

\subsection{Weak joint probabilities and weak purity}

The weak values $A^a(b|\psi)$ and $A^a_w(a|\rho)$ in Eqs.~(\ref{probcom}) and (\ref{comgen}), for the projection operator $A^a$ postselected on measurement result $B=b$, 
are sometimes referred to as  `weak probabilities' \cite{ams1,ams2,ozawaprob,rub}.  However, here we will follow the common practice of identifying the real part of this quantity as 
a weak probability \cite{dressint,weakprob,lund,weston,joh2,rozema}.  More generally, for two arbitrary POVM observables 
$A\equiv\{A_a\}$ and $B\equiv\{B_b\}$ we define the weak conditional  probability of $A=a$ postselected on outcome $B=b$ by
\begin{equation} \label{wcond}  
 p_w(a|b) := {\rm Re}\left\{\frac{\tr{\rho B_b A_a}}{\tr{\rho B_b}}\right\} . \blk
\end{equation}
Note that it is {\it not} assumed that $A_a$ and $B_b$ are rank-1 projection operators, in contrast to $A^a$ and $B^b$ of the previous section.

The weak conditional probability distribution is physically measurable via suitable weak measurements, as is the correpsonding weak {\it joint} probability distribution \cite{dressint,weakprob,lund,weston}
\begin{eqnarray} \label{pwab}
p_w(a,b) := p_w(a|b)\,p(b) =\half\langle A_aB_b + B_bA_a\rangle_\rho ,
\end{eqnarray}
Here $p(b):=\tr{\rho B^b}$ is the probability of measurement outcome $B=b$. 
The weak joint probability distribution may also be recognised as the Margeneau-Hill quasiprobability distribution \cite{mh}, and
satisfies 
\[ \sum_b p_w(a,b) = p(a),~~~\sum_a p_w(a,b) = p(b), \]
just as for classical joint distributions.  However, the weak joint probabilities $p_w(a,b)$ can take anomalous values, lying outside the unit interval $[0,1]$. 

In particular, at least one value of $p_w(a,b)$ is anomalous if and only if
\begin{equation} \label{iff}
\sum_{a,b} |p_w(a,b)| >1, 
\end{equation}
which is equivalent to at least one value being negative. 
Clearly, a sufficient condition for an anomalous value is then that 
\beq 
P_W :=\sum_{a,b} p_w(a,b)^2 
\eeq
is greater than unity, where $P_W$ will be called the weak purity of $p_w(a,b)$ in analogy to classical and quantum purities. Surprisingly, however, it will be shown further below that this condition is 
never satisfied!  That is, while individual weak probabilities can be negative, the weak purity is always bounded by unity, just as for 
classical distributions. This property turns out to be closely related to limits on the incompatibility of quantum observables.

\subsection{Incompatibility}

A natural measure of the incompatibility between two POVM elements $A_a$ and $B_b$ in a quantum state $\rho$ is given by
\beq 
\label{iab}
I(a,b):= \frac{1}{4} |\langle [A_a,B_b]\rangle_\rho|^2 \geq 0 .  
\eeq
Note that $I(a,b)$ vanishes if and only if $[A_a,B_b]\rho=0$.  It follows from a result of Busch \cite{busch} that if one of $A_a$ and $B_b$ is a projector, then $I(a,b)=0$ is equivalent to $p_w(a,b)\geq 0$.  Thus there is a connection between anomalous weak probabilities and incompatibility, which will be sharpened further below.

The incompatibility of two measurements $A$ and $B$ is naturally defined as the total incompatibility of their POVM elements, i.e., 
\beq 
I(A,B) := \sum_{a,b} I(a,b).  
\eeq
Note that this measure is independent of the particular values assigned to the outcomes of $A$ and $B$, and hence characterises incompatibility in an 
invariant manner. The incompatibility ranges between 0 and 1, i.e.,
\beq \label{range}
0\leq I(A,B) \leq 1 . \eeq
The lower bound is is reached if and only and if $[A_a,B_b]\rho=0$ for all $a$ and $b$.  To demonstrate the upper bound, note that 
\begin{align}
I(a,b)&= \left({\rm Im}\left\{ \tr{\rho A_aB_b}\right\}\right)^2 \nn\\
&\leq  \left| \tr{(B_b\rho^{1/2})\,(\rho^{1/2} A_a)}\right|^2 \nn\\
&\leq \tr{\rho (A_a)^2}\,\tr{\rho (B_b)^2}\nn\\
&\leq  \tr{\rho A_a}\,\tr{\rho B_b}\nn\\
&= p(a)\,p(b),
\end{align}
where the Schwarz inequality and  $0\leq A_a,B_b\leq 1$ have been used.  Summing over $a$ and $b$ yields the desired result.  Stronger upper bounds, 
involving the weak purity and the quantum purity,  are given below.

\subsection{Tradeoff relations for purity and incompatibility}

To connect the ideas of the previous two subsections, and also relate them to results of section III, note  that one has the relation
\beq \label{con}
p_w(a,b)^2 + I(a,b) = |\tr{\rho A_a B_b}|^2 . %= |\langle (A_a)^*_w (B_b)_w\rangle|^2.
\eeq
Upper bounding the right hand side yields a number of interesting results which we explore below.

\subsubsection{Pure states and nondegenerate observables}

First, consider the special case of a pure state and nondegenerate (or maximal) observables, i.e., 
\[ \rho=|\psi\rangle\langle\psi|,~~A_a=|a\rangle\langle a|,~~B_b=|b\rangle\langle b|. \]
Here the states $\{|a\rangle\}$ and $\{ |b\rangle\}$ are not assumed to be orthonormal.  Equation (\ref{con}) then simplifies to 
\beq \label{pure}
p_w(a,b)^2 + I(a,b) = |\langle a|b\rangle|^2\, p(a)\,p(b).
\eeq
It may be shown, via Eqs.(\ref{wcond}), (\ref{pwab}) and (\ref{iab}), that this is equivalent to the equality in the complementarity relation Eq.~(\ref{comp}).  Thus, the latter relation may also be interpreted as relating  weak joint probabilities
 and incompatibility. Further, from Eq.~(\ref{pwab}) and recalling $I(a,b)\geq0$, one has the weaker complementarity relation
\beq  \label{compr}
p_w(a|b)\,p_w(b|a) \leq |\langle a|b\rangle|^2 \leq 1, \eeq
relating weak probabilities for a weak measurement of $A$ postselected on a strong measurement of $B$ and vice versa.  Thus,  if the weak probability of $A=a$, 
postselected on $B=b$, is greater than 1, then the converse weak probability must be less than 1.  

Summing Eq.~(\ref{pure}) over $a$ and $b$ yields the tradeoff relation 
\beq \label{overlap}
P_W + I(A,B) \leq c_{AB} := \max_{a,b} |\langle a|b\rangle|^2 \leq 1, \eeq
between the weak purity and the total incompatibility of $A$ and $B$. Note that the maximum overlap of the POVM elements, $c_{AB}$, commonly appears in 
entropic uncertainty relations \cite{ent}.  This relation provides a far stronger upper bound for $I(A,B)$ than does Eq.~(\ref{range}), and will be discussed further below 
in its more general form.

\subsubsection{General tradeoff relations}

For the general case of arbitrary states and observables, one has from the Schwarz inequality that
\begin{align} \nn
|\tr{\rho A_a B_b}|^2 &= |\tr{\rho(A_aB_b)}|^2\\  \nn
& \leq \tr{\rho^2}\,\tr{(A_a)^2(B_b)^2} \\
& \leq \tr{\rho^2}\,\tr{A_a B_b},
\end{align}
where the final inequality makes use of the property $0\leq A_a, B_b\leq 1$.  Substitution into Eq.~(\ref{con}) then gives
\beq \label{tradeoff}
p_w(a,b)^2 + I(a,b)  \leq \tr{\rho^2}\,\tr{A_a B_b} ,
\eeq                            
and summation over $a$ and $b$ yields the tradeoff relation
\beq \label{diff}
P_W + I(A,B) \leq \tr{\rho^2}=: P_Q,
\eeq
where $P_Q$ denotes the quantum purity.  It also follows, recalling $I(A,B)\geq0$, that 
\beq P_W \leq P_Q,  \eeq
i.e., the weak purity can never exceed the quantum purity.

The tradeoff relation in Eq.~(\ref{diff}) generalises Eq.~(\ref{overlap}) to all states and observables, and has several physical implications.  
First, as discussed  in the previous section, the weak purity is never greater 
than the classical maximum value of 1,
even when some of the weak probabilities are negative, thus restricting the degree to which the weak probabilities can take anomalous values.  Second, the greater 
the incompatibility of $A$ and $B$, the smaller the weak purity, and vice versa.  Third, the incompatibility of $A$ and $B$ is upper-bounded by the difference
between the quantum purity and the weak purity. This difference can, therefore, be considered a resource for incompatibility, analogous to the manner in which 
the difference between a quantum purity and a classical purity acts as a resource for quantum coherence \cite{coh}.

Finally, note that an alternative application of the Schwarz inequality in the general case yields
\begin{align} \nn
|\tr{\rho A_a B_b}|^2 &= \left|\tr{(B_b \rho^{1/2})(\rho^{1/2} A_a)}\right|^2\\
& \leq \langle (B_b)^2\rangle_\rho\,\langle (A_a)^2\rangle_\rho\\
& \leq \langle A_a\rangle_\rho\,\langle B_b\rangle_\rho,
\end{align}
Substitution into Eq.~(\ref{con}) then gives
\beq  \label{first}
p_w(a,b)^2 + I(a,b) \leq p(a)\, p(b), 
\eeq
leading immediately to the generalisation, 
\beq p_w(a|b)\,p_w(b|a) \leq 1 \eeq
of the complementarity relation in Eq.~(\ref{compr}), to arbitrary density operators and observables.  

\subsection{Weak probabilities vs strong probabilities}

Weak measurements are considered to be `weak' because they involve a weak interaction between the system and an apparatus.  Here we demonstrate a connection between the statistics of weak and strong measurements that 
 yields a different sense of `weakness' for weak measurements: anomalous values of weak joint probabilities are restricted by the values of corresponding strong joint probabilities.
 
 In particular, taking $A$ and $B$ to be 
projective measurements for simplicity,  the Schwarz inequality yields
\begin{align} \nn
|\tr{\rho A_a B_b}|^2 &= \left|\tr{\rho^{1/2}(\rho^{1/2}A_aB_b)}\right|^2\\ \nn
& \leq \tr{A_a\rho A_a B_b} =: p_s(a,b).
\end{align}
The quantity $p_s(a,b)$ may be recognised as the joint probability of outcomes $A=a$ and $B=b$, for the scenario in which a {\it strong} measurement of $A$ is followed by a strong
measurement of $B$.  Substitution into Eq.~(\ref{con}) then gives the fascinating connection 
\beq \label{strong}
p_w(a,b)^2 + I(a,b) \leq p_s(a,b),
\eeq
linking weak probabilities, strong probabilities, and incompatibility.  

For example, we have from Eq.~(\ref{strong})  that
\beq |p_w(a,b)| \leq \sqrt{p_s(a,b)}\leq 1.  \eeq
Hence, individual weak joint probabilities are bounded 
in modulus by the square root of a classical probability.   

As a second example, note from Eq.~(\ref{strong}) that 
\beq 
I(a,b) \leq p_s(a,b),
\eeq
i.e., the statistics of two successive strong projective measurements bound the incompatibility of the corresponding measurement outcomes.
Generalizations of these results can obtained for non-projective measurements, but will not be considered here.

\section{Conclusions}

We have shown that the products of weak values of quantum observables have many physical applications.  For example, a product representation formula can be used to recover quantum averages of products of observables from experimentally determined weak values of the observables; to show that the standard 
Heisenberg uncertainty relation is equivalent to a classical uncertainty relation for complex random variables; to derive strong uncertainty relations for
pairs of unitary operators; and to obtain an interpretation of weak values as optimal estimates. Further, for two complementary weak measurement scenarios, in which 
the weakly and strongly measured observables are interchanged, there is a 
complementarity relation in the form of an upper bound on the product of the corresponding weak values. 
Finally, general trade-off relations have been  obtained between 
weak purities, quantum purities and the degree of incompatibility of two observables, and also between the corresponding weak and strong joint probability distributions, 
which quantify the extent to which weak probabilities can take anomalous values. We hope that our results will open up new ways of thinking about uncertainty and 
complementarity relations using products of weak values. In future, it may be worth exploring if the  hidden variable model and product representation 
formula, based on complex weak values, can reproduce the nonlocal correlations of entangled states.

\vskip 1cm

\noindent
{\bf Acknowledgement:}

 MH is  supported by the ARC Centre of Excellence CE110001027. 
J. Wu and A. K.  Pati are supported by National Natural Science Foundation of China (11171301 and 11571307) and the Doctoral Programs Foundation of Ministry of Education of China (J20130061).
\blk


\newpage

\begin{thebibliography}{999}

\bibitem{aha} Y. Aharonov, D. Z. Albert, and L. Vaidman, Phys. Rev. Lett. {\bf 60}, 1351 (1988). 	

\bibitem{av} Y. Aharonov and L. Vaidman, Lect. Notes Phys. {\bf734}, 399 (2008).

\bibitem{average}  The fact that weak values arise experimentally as {\it averages} of certain measurements, rather than as measurement outcomes {\it per se}, suggests that `weak averages' is a more appropriate terminology. However, we conform with existing literature here.

\bibitem{akp} A. K. Pati and J. Wu, arXiv:1410.5221 (2014).

\bibitem{sw1} S. Wu and Y. Li, Phys. Rev. A {\bf83}, 052106 (2011).

\bibitem{ADL} A. DiLorenzo, Phys. Rev. A {\bf85}, 032106 (2012).

\bibitem{sw2} S. Pang, S. Wu, and Z.B. Chen, Phys. Rev. A {\bf86}, 022112 (2012).

\bibitem{nori} A. G. Kofman, S. Ashhab, and F. Nori, Physics Reports {\bf520}, 43–133 (2012).

\bibitem{shik} Y. Shikano, and A. Hosoya, J. Phys. A: Math. Theor. {\bf43}, 025304 (2010).

\bibitem{rub} K. Fukuda, J. Lee and I. Tsutsui, Eprint arXiv:1602.08872 [quant-ph] (2016).

\bibitem{eric} E. Sj{\"o}qvist, Phys. Lett. A {\bf359}, 187 (2006).

\bibitem{sandu} Y. Aharonov, S. Popescu, and J. Tollaksen, Phys. Today {\bf 63}, 27 (2010).

\bibitem{cho} A. Cho, Science {\bf 333}, 690 (2011).

\bibitem{wise1} H. M. Wiseman, Phys. Lett. A {\bf 311}, 285 (2003).

\bibitem{mir}  R. Mir, J. S. Lundeen, M. W. Mitchell, A. M. Steinberg, J. L. Garretson and H. M. Wiseman, New J. Phys. {\bf 9}, 287 (2007).

\bibitem{js} J. S.  Lundeen and A. M.  Steinberg, Phys. Rev. Lett. {\bf 102}, 020404 (2009).

\bibitem{ams1} A. M. Steinberg, Phys. Rev. Lett. {\bf74}, 2405 (1995).

\bibitem{ams2} A. M. Steinberg, Phys. Rev. A {\bf52}, 32 (1995).

\bibitem{pdav} P. C. W. Davies, Phys. Rev. A {\bf79}, 032103 (2009).

\bibitem{jsl1} J. S. Lundeen, B. Sutherland, A. Patel, C. Stewart, and C. Bamber,
Nature {\bf474}, 188 (2011).

\bibitem{hof} H. F. Hofmann, Phys. Rev. A {\bf 83}, 022106 (2011).

\bibitem{jd} J. Dressel, M. Malik, F. M. Miatto, A. N. Jordan, R. W. Boyd,
Rev. Mod. Phys. {\bf86}, 307 (2014).

\bibitem{geno} F. Piacentini, M. P. Levi, A. Avella, E. Cohen, R. Lussana, F. Villa, A. Tosi, F. Zappa, M. Gramegna, G. Brida, I. P. Degiovanni, and M. Genovese, arXiv.1508.03220v1 [quant-ph] (2015).

\bibitem{joh1} L. M. Johansen, Phys. Lett. A {\bf 322}, 298 (2004).

\bibitem{hall} M. J. W Hall, Phys. Rev. A {\bf 69}, 052113 (2004).

\bibitem{joh2} L. M. Johansen, Phys. Lett. A {\bf 366}, 374 (2007).

\bibitem{joz} R. Jozsa, Phys. Rev. A {\bf 76}, 044103 (2007).

\bibitem{dress} J. Dressel and A. N. Jordan, Phys. Rev. A {\bf 85}, 012107 (2012).

\bibitem{dressint} J. Dressel, Phys. Rev. A {\bf 91}, 032116 (2015).

\bibitem{inter} Y.-X. Zhang, S. Wu and Z.-B. Chen, Phys. Rev. A {\bf 93},  032128 (2016).
\blk

\bibitem{cat1} Y. Aharonov, S. Popescu, D. Rohrlich and P. Skrzypczyk, New J. Phys. {\bf 15}, 113015 (2013).

\bibitem{cat2} A. Matzkin and A. K. Pan, J. Phys. A: Math. Theor. {\bf 46}, 315307 (2013).

\bibitem{cat3} T. Denkmayr,	H. Geppert,	S. Sponar,	H Lemmel,	A. Matzkin, J. Tollaksen and Y. Hasegawa, Nature Commun. {\bf 5}, 4492 (2014).	

\bibitem{weakprob} J. L. Garretson, H. M. Wiseman, D. T. Pope and D. T. Pegg, J. Opt. B {\bf 6}, S506 (2004).

\bibitem{mo1} M. Ozawa, Phys. Rev. A {\bf 67}, 042105 (2003). 	
\bibitem{weston} M. M. Weston, M. J. W. Hall, M. S. Palsson, H. M. Wiseman and G. J. Pryde, Phys. Rev. Lett. {\bf 110}, 220402 (2013).

\bibitem{bra} C. Branciard, Proc. Natl. Acad. Sci. USA {\bf 110}, 6742 (2013). 	
\bibitem{cbra} C. Branciard, Phys. Rev. A {\bf 89}, 022124 (2014). 	

\bibitem{ozmixed} M. Ozawa, Eprint  arXiv:1404.3388v1 [quant-ph] (2014).

\bibitem{density}  Such hidden variable models, and the product representation formula, may be formally extended to the case of a quantum state described by density operator $\rho$. In this case one requires two complete POVMs, $M\equiv \{|m\rangle\langle m|\}$ and $N\equiv \{|n\rangle\langle n|\}$, and each operator $A$ is assigned the value $A(m,n|\rho):=\langle m|A\rho^{1/2}|n\rangle / 
\langle m|\rho^{1/2}|n\rangle$ with probability $p(m,n|\rho):=|\langle m|\rho^{1/2}|n\rangle|^2$. Note this reduces to the usual weak value in the case of pure states, but more generally is different to the standard definition in Eq.~(\ref{weakrho}). \blk

\bibitem{sig1} F. D. Neeser and J. L. Massey, IEEE Trans. Inf. Theory {\bf 39}, 1293 (1993).

\bibitem{sig2} B. Hajek, {\it Random Processes for Engineers} (Cambridge University Press, UK, 2015), chaps.~7,8.

\bibitem{qfunction} H.M. Wiseman and G. J. Milburn, {\it Quantum Measurement and Control} (Cambridge University Press, UK, 2010), App.~A.5.


\bibitem{heis} W. Heisenberg, 
Z. Phys. {\bf 43},  172 (1927).


\bibitem{ken} E. H. Kennard, Z. Phys. {\bf 44}, 326 (1927).


\bibitem{rob} H. P. Robertson,

Phys.  Rev. {\bf 34}, 163 (1929).

\bibitem{schr} E. Schr\"odinger, Ber. Kgl. Akad. Wiss. Berlin {\bf 19}, 296 (1930).


\bibitem{lorenzo} L. Maccone and A. K. Pati,  Phys. Rev. Lett. {\bf 113}, 260401 (2014).


\bibitem{levy} J.-M. L\'evy-Leblond, Ann. Phys. NY {\bf 101}, 319 (1976).

\bibitem{pati} A. K. Pati, U. Singh and U. Sinha, Phys. Rev. A {\bf 92}, 052120  (2015).

\bibitem{massar} S. Massar and P. Spindel, Phys. Rev. Lett.{\bf 100}, 190401 (2008).

\bibitem{patiun} S. Bagchi and A. K. Pati, Eprint arXiv:1511.04730v1 (2015).

\bibitem{rudnicki} L. Rudnicki, D. S. Tasca, and S. P. Walborn, Phys. Rev. A {\bf 93}, 022109  (2016).
 
\bibitem{barg} V. Bargmann, J. Math. Phys. {\bf 5}, 862 (1964).

\bibitem{msd} Advantages and disadvantages of such `noise operator' measures has been reviewed recently in M. A. Appleby, Eprint arXiv:1602.09002v1 (2016).

 
\bibitem{bohr} N. Bohr, Phys. Rev. {\bf 48}, 696 (1935).


\bibitem{ozawaprob} M. Ozawa, AIP Conf. Proc. {\bf 1363}, 53 (2011).

\bibitem{lund} A. P. Lund and H. M. Wiseman, New J. Phys. {\bf 12}, 093011
(2010).

\bibitem{rozema} L. A. Rozema, A. Darabi, D. H. Mahler, A. Hayat, Y. Soudagar, and A. M. Steinberg, Phys. Rev. Lett. {\bf 109}, 100404, (2012). 

\bibitem{mh} H. Margenau and R. N. Hill, Prog. Theor. Phys. {\bf 26}, 722
(1961).

\bibitem{busch} P. Busch and T. Heinosaari, Quantum Inf. Comp. {\bf 8}, 797 (2008).

\bibitem{ent} P. J. Coles, M. Berta, M. Tomamichel and S. Wehner, Eprint  arXiv:1511.04857v1 (2015).


\bibitem{coh} S. Cheng and M. J. W. Hall, Phys. Rev. A {\bf 92}, 042101  (2015).




\end{thebibliography}
\end{document}